# High Performance N-Type Carbon Nanotube Field-Effect Transistors with Chemically Doped Contacts

Ali Javey[1], Ryan Tu[1], Damon Farmer[2], Jing Guo[3], Roy Gordon[2], and Hongjie Dai[1*]

[1] Department of Chemistry and Laboratory for Advanced Materials, Stanford University, Stanford, CA 94305, USA

[2] Department of Chemistry and Chemical Biology, Division of Engineering and Applied Sciences, Harvard University, Cambridge, MA 02138, USA

[3] Department of Electrical and Computer Engineering, Univ. of Florida, Gainesville, FL, 32608

**Abstract.** Short channel (~80 nm) n-type single-walled carbon nanotube (SWNT) field-effect transistors (FETs) with potassium (K) doped source and drain regions and high-$\kappa$ gate dielectrics (ALD $HfO_2$) are obtained. For nanotubes with diameter ~ 1.6 nm and bandgap ~ 0.55 eV, we obtain n-MOSFET-like devices exhibiting high on-currents due to chemically suppressed Schottky barriers at the contacts, subthreshold swing of 70mV/decade, negligible ambipolar conduction and high on/off ratios up to $10^6$ at a bias voltage of 0.5V. The results compare favorably with the state-of-the-art silicon n-MOSFETs and demonstrate the potential of SWNTs for future complementary electronics. The effects of doping level on the electrical characteristics of the nanotube devices are discussed.

* Email: hdai@stanford.edu

Single-walled carbon nanotubes are promising for future high performance electronics such as field effect transistors owing to their various unique properties including ballistic transport with relatively long mean free paths and high compatibility with high-κ dielectrics imparted by their unique chemical bonding and surface stability.[1-12] While much has been done to achieve high performance p-type nanotube FETs through contact optimization, dielectric integration and lateral scaling, progress on n-FETs has been slow partly due to the difficulty in affording low Schottky (SB) contacts for high on-states and at the same time achieving high on/off ratios with small diameter (or large bandgap) tubes.[7] Here, by invoking chemical doping, high-κ dielectrics and new device design, we demonstrate n-type SWNT FETs with performance matching or approaching the best p-type nanotube FETs and surpassing the state-of-the-art Si n-MOSFET.

SWNT synthesis,[13] atomic layer deposition (ALD) of $HfO_2$[10,14,15] ($t_{ox}$=8nm) and details of device fabrication are similar to those described previously. In Fig.1, we first show the electrical properties of a back-gated ($t_{SiO2}$=10nm, Fig. 1a) semiconducting SWNT (d~1.4 to 1.5 nm, channel length L~150 nm between Pd source/drain S/D) device before and after K-doping (details of doping described previously[16-18]). The as-made device is a p-type FET with $I_{on}$~ 5μA and linear conductance of $G_{on}$~0.3 $e^2$/h (Fig. 1b). $I_{on}$ is lower than the expected[1,19] 20-25 μA per tube as limited by the existence of a SB[4] (height ~0.1 eV, width ~ $t_{SiO2}$) between Pd[1] and the d=1.4-1.5nm tube. After heavily n-doping the device with potassium in vacuum, $I_{on}$ and $G_{on}$ increase to over 20 μA and $e^2$/h (Fig. 1b) respectively, suggesting high metal-semiconductor contact transparency ($T_{MS}$ ~ 1) and quasi ballistic transport within the nanotube. The current-voltage ($I_{ds}$-$V_{ds}$) characteristics of the device become largely gate independent, corresponding to a near-metallic $n^{++}$ state of the SWNT (Fermi level well within the conduction band) due to heavy electron-donation by K. This result clearly shows that n-doping of the semiconducting SWNT is highly effective in suppressing SBs and increasing the electron transmission probability $T_{MS}$ to the conduction band of SWNTs at the Pd-tube contacts.

Having established that chemical doping can afford high on-current injection into the n-channel of a SWNT, we then moved onto constructing n-type SWNT FETs with $n^+$-i-$n^+$ structures (with heavily doped S/D regions, Fig. 2a) similar in principle to conventional n-MOSFETs.[20] We patterned top-gate stack (gate length $L_g$~80 nm) on individual SWNTs on $p^+$Si/SiO$_2$ (~500 nm) substrates without overlapping the Pd (15 nm thick) metal S/D (distance L~250 nm). The top-gate stack consisted of a ~ 8 nm thick HfO$_2$ dielectric layer and ~0.5/15 nm Ti/Pd gate metal atop formed by an ALD and liftoff technique.[10,14,15] The nanotube segments outside the gate stack are fully exposed for K vapor doping to form $n^+$ regions (Fig.2a). ALD of HfO$_2$ on SWNTs provides excellent electrostatic modulation of the channel conductance without degrading the transport property of the 1D nanotube channels.[3,9,10] This is another key element in affording high performance n-type SWNT FETs.

Prior to K doping, the devices operated as p-MOSFETs (Fig. 2b blue curve) when the two ends of the tube were electrostatically hole-doped by a back-gate.[21] Upon exposure to K vapor in vacuum, the S/D regions became n-doped while the top-gated channel regions remained intrinsic due to blocking of K by the gate stack. This afforded $n^+$-i-$n^+$ n-type nanotube MOSFETs. The electrical properties of a SWNT (d~1.6-1.7 nm; $E_g$ ~0.55eV) MOSFET before (p-type) and after K-doping (n-type) in vacuum are shown in Fig. 2b-2c. The n-FET (K-doped S/D) and p-FET (electrostatically doped S/D) showed near-symmetrical characteristics with similar on-currents $I_{on} \sim 8$ µA at $V_{ds}$=0.5 (Fig. 2c). The transconductance were $(dI_{ds}/dV_{gs})_{max} \sim 20$ µS and ~10 µS for n- and p-FETs respectively. Both devices exhibited excellent switching characteristics with subthreshold swings $S=dI_{ds}/dV_{gs} \sim 70$-80 mV/decade (Fig. 2b), near the theoretical limit of $S \sim 60$ mV/decade. Note that at $V_{ds}$=0.5V, a high $I_{on}/I_{off} \sim 10^6$ was achieved for the nanotube n-MOSFET with no significant ambipolar p-channel conduction (Fig. 2b red curve). These characteristics are the best reported to-date for n-type nanotube FETs enabled by the MOSFET geometry[6,22] with chemically doped S/D, high-$\kappa$ dielectrics and transparent metal-tube contacts. In such a MOSFET-like geometry the gate electric fields result in bulk switching of the nanotube directly under the gate-stack with little effect to the Schottky barriers at the metal-tube junctions.

We next investigated the effects of the contacts doping level on the electrical characteristics of our nanotube n-FETs. The doping level was varied by adjusting the exposure time of the devices to K atoms. Fig. 3a (dashed curve) shows the switching properties for the same device in Fig.2 but at a higher degree of n-doping of the S/D contacts. The on-current of the device increased from ~ 8µA to 15 µA at $V_{ds}$=0.5 V (Fig. 3b), attributed to further enhanced transparency at the Pd-tube junctions and lower series resistance in the $n^+$ nanotube segments. The on-current increase was, however, accompanied by a more obvious ambipolar p-channel conduction, an increase in the minimum leakage current ($I_{min}$) and a reduction of $I_{on}/ I_{min}$ from $10^6$ to $10^4$ (Fig.3a dashed curve). The enhanced $I_{min}$ is attributed to the thinning of the band-to-band tunneling barriers (resulting in an increase in the transmission probability $T_{bb}$) and reduction of the activation energy barriers for the thermionic emission of electrons at the doped-contact/channel junctions (Fig. 3d). Beside $I_{on}$ and $I_{min}$, the ambipolar hole leakage current at large negative voltages is also enhanced at the higher doping level, once again, due to the decrease in the band-to-band tunneling barriers.

The results above suggest that for d~1.6nm and bandgap $E_g$~0.55eV SWNTs, a moderate doping level for the $n^+$ S/D regions is optimum for high performance nanotube n-MOSFET with $I_{on}$ of ~8µA, S~70mV/decade, small leakage current ($I_{on}/I_{off}$~$10^6$) and little ambipolar conduction. Comparing to our best p-type SWNT FETs,[10] the current n-FET is better in lower off-state and less ambipolar conduction (due to the MOSFET geometry[6,22] and smaller tube diameter) but lower in on-current due to the series resistance in the S/D nanotube segments and non-ideal n-type contacts at this doping level. Careful attention must be paid in controlling the chemically doped S/D contacts for carbon nanotube MOSFETs in order to balance $I_{on}$ and $I_{on}/I_{off}$. This is in contrast to the Si MOSFETs where band-to-band tunneling is not significant and the leakage currents are mostly independent of the S/D doping profiles, at least in the ~100 nm length scales. The enhanced band-to-band tunneling in nanotube devices are attributed to smaller energy band gaps, smaller effective carrier mass, and symmetric conduction and valance band states. Potentially, larger bandgap (smaller diameter) nanotubes can be integrated as the active device components in order to allow for lower band-to-band leakage currents even at very high contact doping levels, enabling higher on-state currents and lower leakage currents.

We note that the n-channel on-state conductance of the nanotube FET with highly n-doped contacts exhibits little temperature dependence from 300 K to 10 K and oscillations in $I_{ds}$-$V_{gs}$ at low temperatures due to quantum interference effects (Fig. 3c). This confirms the absence of thermionic current to the conduction band of SWNT at the Pd/n$^+$-SWNT contacts for the heavy K-doping case (with high $I_{on}$~20 µA). The SB height between a conventional semiconductor (e.g., Si) and a metal is relatively insensitive to doping level in the semiconductor due to Fermi level pinning by surface states. Ohmic contacts to a conventional semiconductor are obtained by heavy doping to afford ultra-thin SB barrier width transparent to tunneling.[20] Since Fermi level pinning is small or nearly absent in SWNTs, we suggest that near-transparent n-type contacts formed by chemical doping are mainly a result of reduction of SB height at the tube-metal junctions. It is also possible that thinning of SB width is involved, though the precise degree of pinning and barrier thinning are currently unknown and require further investigation.

To further assess the performance of our n-type nanotube FETs, we compare their on-currents per unit width (a normalization factor of 2d) with the state-of-the-art Si n-MOSFETs at the same power supply voltage and on-off ratios.[24] On-current is an important figure of merit for transistor performance since it is linearly proportional to the device speed. At a power supply voltage of $V_{dd}$=$V_{ds}$=$V_{gs}$(on)-$V_{gs}$(off)=0.5 V, we show in Fig. 4 the on-state current density ($I_{on}$/2d) vs. $I_{on}$/ $I_{off}$ for our n-FETs and for that of a 100 nm node Si n-MOSFET with gate length $L_g$~50 nm (the on and off states for nanotube FETs were determined using a method recently described[25]). One sees that for $I_{on}$/$I_{off}$~10$^2$, the more heavily doped nanotube FET can deliver twice as much on-current as the moderately doped device (~2000 µA/µm vs. ~1100 µA/µm). For $I_{on}$/ $I_{off}$~10$^4$, the latter can deliver ~ 300 µA/µm on-current and the $I_{on}$/ $I_{off}$~10$^4$ is unattainable by the more heavily doped device. The on-current for the Si n-MOSFET is lower than both the heavily and moderately doped S/D CNTFETs at all $I_{on}/I_{off}$ values, with $I_{on}$~320 and 120 $\mu A/\mu m$ at $I_{on}/I_{off}$~$10^2$ and $10^4$ respectively. Note that the comparison here should be taken as specific to the method of current normalization by a channel width of $2d$ for the nanotube FET. Devices with well-spaced and packed SWNTs in parallel will need to be obtained for the ultimate comparison with Si MOSFETs.

To summarize, high performance short channel (~80 nm) n-type nanotube MOSFETs with chemically doped source and drain regions and high-κ gate dielectrics are obtained. For nanotubes with diameter ~1.6 nm and bandgap ~ 0.55 eV, we obtain n-MOSFET exhibiting high on-currents, subthreshold swings of 70mV/decade, small ambipolar conduction and high on/off ratios up to $10^6$ at a drive voltage of 0.5V. The results compare favorably with the state-of-the-art silicon n-MOSFETs and demonstrate the potential of SWNTs for future complementary electronics.

**Acknowledgements:**

We are grateful to Prof. Mark Lundstrom for insightful and stimulating discussions. We thank Prof. D. Antoniadis and O. Nayfeh of MIT for providing the Si MOSFET data in Figure 4. This work was supported by Intel, MARCO MSD Focus Center, NSF Network for Computational Nanotechnology, a SRC Peter Verhofstadt Graduate Fellowship (A. J.), and a NSF Graduate Fellowship (R.T.).

# Figure Captions

**Figure 1**. N-doping of nanotubes by K vapor. (a) AFM image of a nanotube device (top panel) and schematic drawing of K doping of the device (bottom panel). The K vapor was generated by applying a current across an alkaline metal dispenser[16] for a few minutes (~1-10 min, depending on the desired doping level). (b) $I_{ds}$-$V_{ds}$ characteristics of a device before and after K-doping at two gate voltages $V_{gs}$=-3V (red) and 0V (green). Heavy n-doping of the nanotube along its entire length resulted in highly conducting near-metallic behavior with no significant gate dependence. All measurements were conducted in vacuum (~$10^{-5}$ torr). The chemical doping effects were found to reverse upon exposure to ambient environment due to the reactivity of K atoms. Potentially, passivation methods or air-stable dopants could be used to avoid air sensitivity.

**Figure 2**. A high performance nanotube n-MOSFET. (a) AFM image and schematic of a nanotube MOSFET. Note that two steps of electron beam lithography with an over-layer positioning accuracy of ~ 50 nm were used to form Pd S/D and gate stack respectively. (b) $I_{ds}$-$V_{gs}$ curves for a nanotube (d~1.6-1.7 nm) MOSFET before (blue) and after (red) K-doping. Before K-doping for the p-FET, a back gate voltage of -15 V was applied to obtain electrostatically p-doped contacts. For the n-FET, the back gate was grounded during the measurements after the contacts were chemically doped by K. (c) $I_{ds}$-$V_{ds}$ output characteristics of the device before (p-FET, blue) and after (n-FET, red) K-doping. From low to high current curves, the top-gate voltages are $V_{gs}$=0.2 to –1 V in 0.2 V steps for the p-FET and $V_{gs}$=-0.2 to 0.6 V in 0.2 V steps for the n-FET.

**Figure 3**. Dependence of nanotube FET transport properties on the doping level of contacts. (a) $I_{ds}$-$V_{gs}$ characteristics of the nanotube n-FET in Fig. 2 with high (dashed line) and moderate (solid line) doping levels. The back gate is grounded in all measurements. (b) $I_{ds}$-$V_{ds}$ curves for the high-doping n-FET with $V_{gs}$=-0.2 to 1V in 0.2 V steps. (c) $I_{ds}$-$V_{gs}$ curves of the high-doping n-FET recorded at 300 K and 10 K respectively. (d) A band diagram for the off-state of the nanotube n-MOSFET. Large contact doping results in higher on-currents (increased metal-semiconductor transmission probabilities $T_{ms}$ and reduced series resistance of the semiconductor contacts), but also higher leakage and more ambipolar conduction (increased band to band transmission probability $T_{bb}$).

**Figure 4**. Nanotube and Si n-MOSFETs performance comparison. Current density (current normalized by 2d for nanotubes) as a function of $I_{on}/I_{off}$ for the nanotube device with two S/D doping levels and a 100 nm node state-of-the-art Si nMOSFET[26] at $V_{dd}$=$V_{ds}$=$V_{gs}$(on)-$V_{gs}$(off)=0.5 V. The Sin MOSFET has a channel length of L~70 nm and a $t_{ox}$~1.5 nm thick thermally grown $SiO_2$ as gate dielectric (Ref. 26 and personal communication with D. A. Antoniadis).

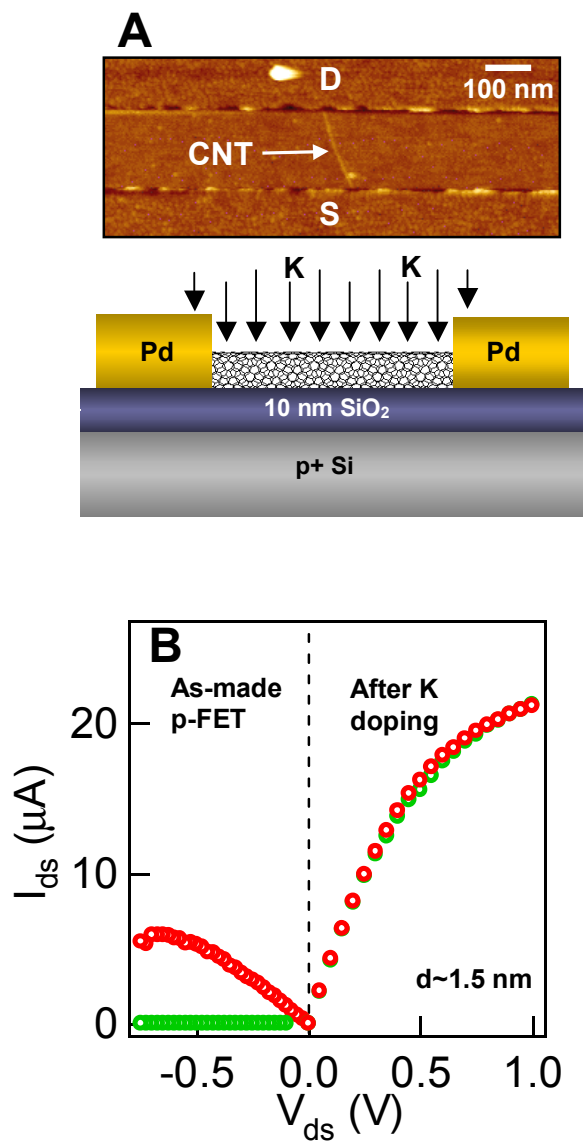

Figure 1

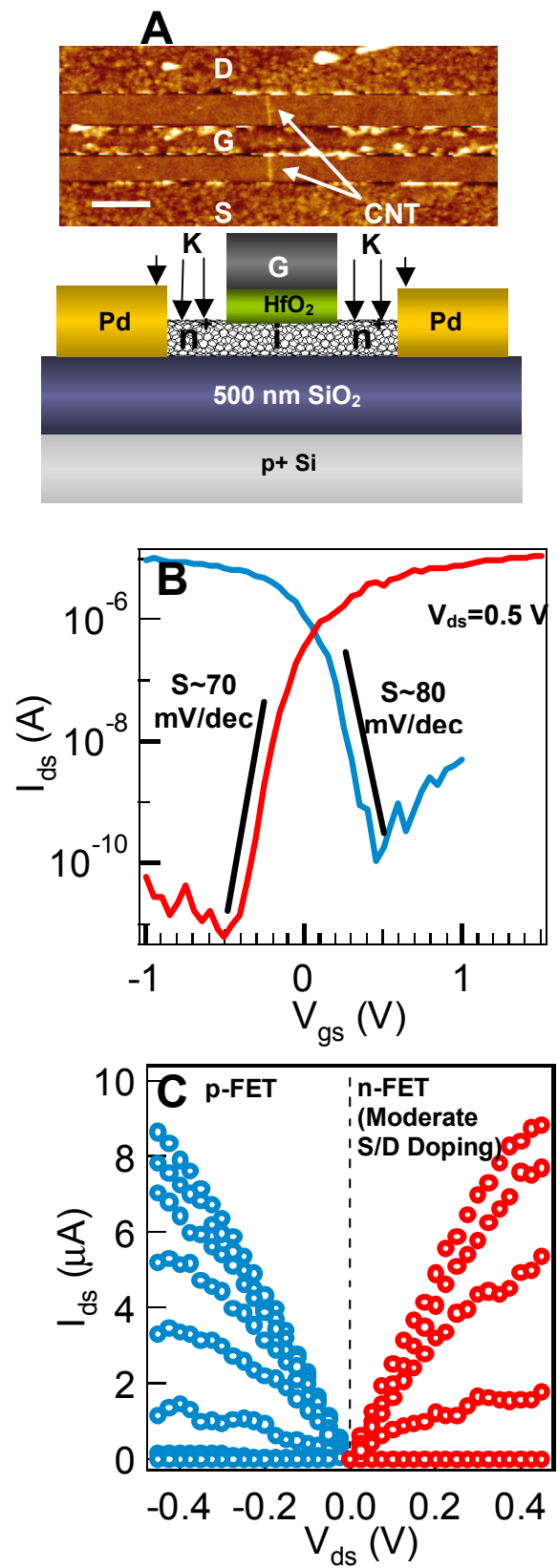

Figure 2

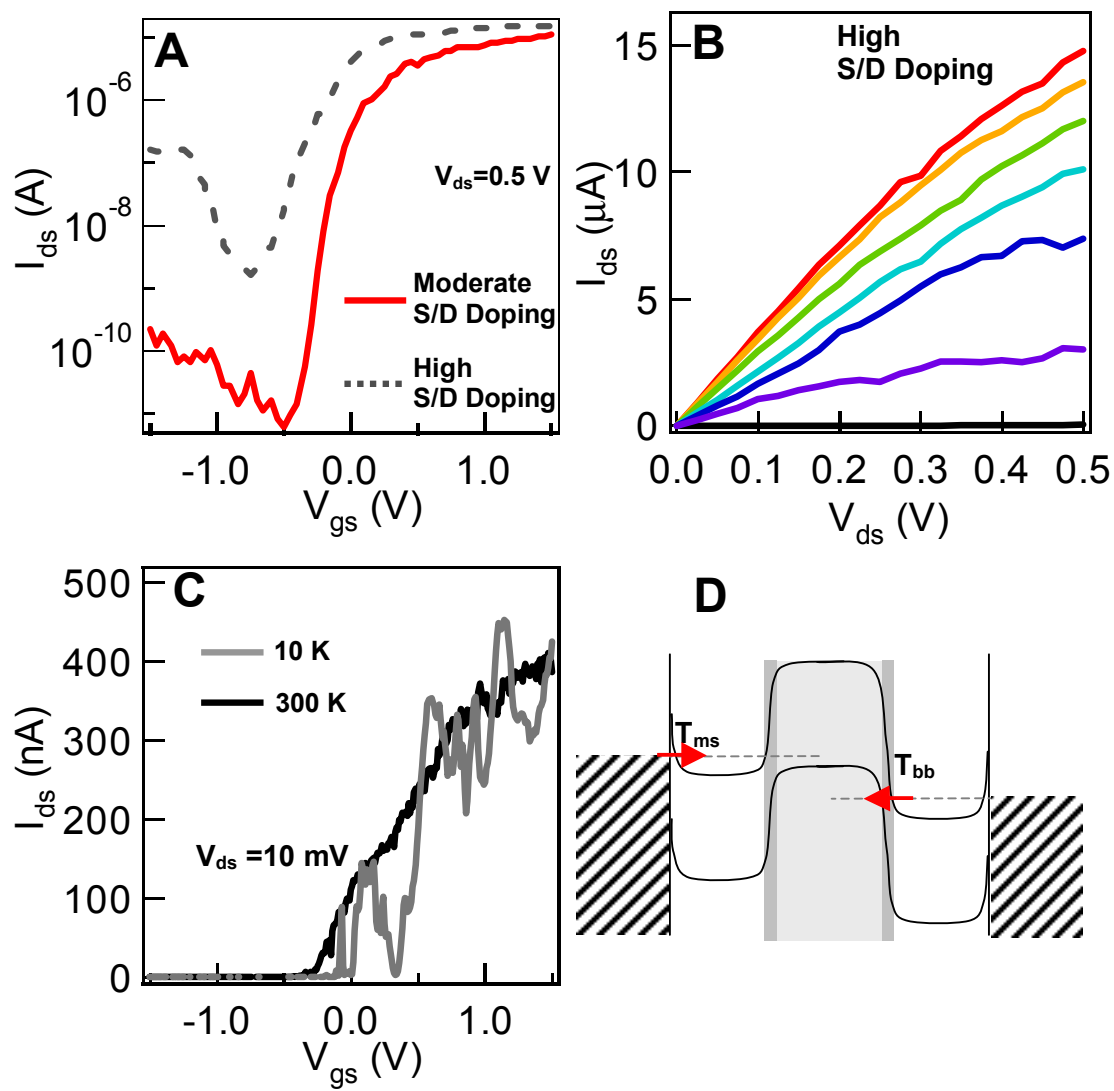

Figure 3

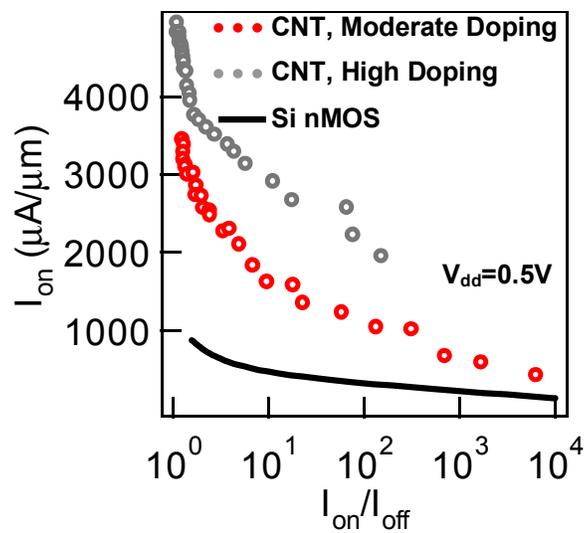

Figure 4

**Table of Contents**

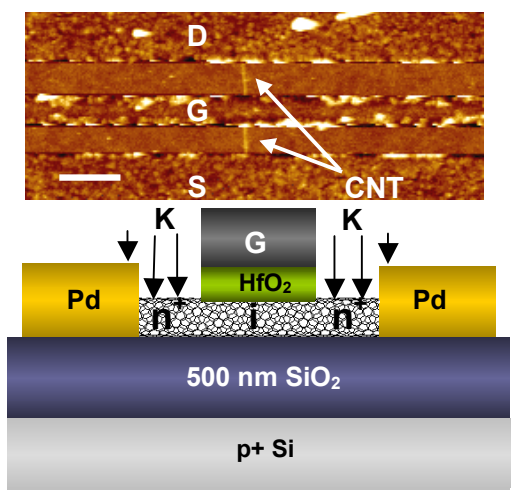

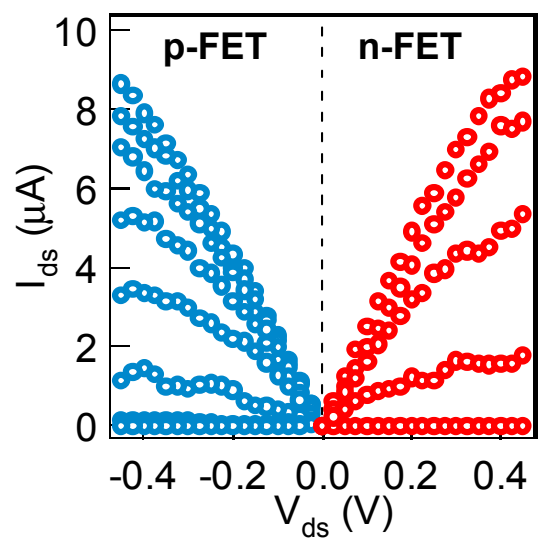